\documentclass[twocolumn,amsmath,amssymb,superscriptaddress]{revtex4}

\begin{document}
\title{Testing Born-Infeld electrodynamics in waveguides}

\author{Rafael Ferraro}
\email{ferraro@iafe.uba.ar}
\thanks{Member of Carrera del Investigador Cient\'{\i}fico (CONICET,
Argentina)}

\affiliation{Instituto de  Astronom\'\i a y F\'\i sica del
Espacio, Casilla de Correo 67, Sucursal 28, 1428 Buenos Aires,
Argentina}\affiliation{Departamento de F\'\i sica, Facultad de
Ciencias Exactas y Naturales, Universidad de Buenos Aires, Ciudad
Universitaria, Pabell\'on I, 1428 Buenos Aires, Argentina.}


\begin{abstract}
Waveguides can be employed to test non-linear effects in
electrodynamics. We solve Born-Infeld equations for TE waves in a
rectangular waveguide. We show that the energy velocity acquires a
dependence on the amplitude, and harmonic components appear as a
consequence of the non-linear behavior.
\end{abstract}


\maketitle

Born-Infeld electrodynamics \cite{B,BI2,BI3,BI4} is a non-linear
theory that modifies Maxwell electromagnetism in the regime of
strong field, which was originally aimed to render finite the
self-energy of a point-like charge. The theory was not
sufficiently appreciated in its epoch because the attention of the
scientific community concentrated mainly on the newborn quantum
electrodynamics. However the interest in Born-Infeld theory
renewed in the last decades because Born-Infeld-like features
emerge in the low energy limit of string theories \cite{Frad,
Abou,Leigh,Met,Tsey,Tsey2}. Born-Infeld theory introduces a new
fundamental constant $b$, with dimensions of electromagnetic
field, which fix the field scale for the transition from Maxwell
weak field regime to Born-Infeld strong field regime. Maxwell and
Born-Infeld theories have proved to be the sole theories for a
massless spin 1 field having causal propagation \cite{Pleb, Deser}
and absence of birefringence \cite{Boillat,Novello} (however
birefringence can exist in theories describing interactions
between Born-Infeld and scalar fields, as in the case of extended
Kaluza-Klein theories \cite{kerner, gibbons}). Some Abelian and
non-Abelian generalizations of Born-Infeld theory has been
mentioned as leading to a non-trivial Galilean limit
$c\rightarrow\infty$ \cite{goldin}.

Born-Infeld plane waves do not differ from Maxwell plane waves;
however, when waves propagate in the presence of background fields
or boundaries then Born-Infeld waves depart from Maxwell ones
owing to non-linear effects. In fact, if background fields are
present then the propagation velocity becomes smaller than $c$\
\cite {Pleb,5pleb,Boillat} (see Ref.\cite{gibbons} for an
equivalent effect in string theory). Moreover, the interaction
with the background field causes a typically anisotropic effect
since the Poynting vector does not share the wave propagation
direction \cite{Aiello}. This kind of effects has not been
observed yet, which means that constant $b$, if it exists, has a
very large value \cite{Jack}. In this work we will study the
propagation of Born-Infeld fields in waveguides to show the
consequences of Born-Infeld non-linearity in a context susceptible
of experimental verification.

Born-Infeld field equations for the antisymmetric field tensor
$F_{\mu \nu }$\ are
\begin{equation}
\partial _{\nu }F_{\lambda \mu }+\partial _{\mu }F_{\nu \lambda }+\partial
_{\lambda }F_{\mu \nu }\ =\ 0\ ,  \label{Feq}
\end{equation}
\begin{equation}
\partial _{\nu }\mathcal{F}^{\mu \nu }=\ 0\ ,  \label{Fcurseq}
\end{equation}
where
\begin{equation}
\mathcal{F}_{\mu \nu }=\frac{F_{\mu \nu }-\frac{P}{b^{2}}^{\ast }\!\!F_{\mu
\nu }}{\sqrt{1+\frac{2S}{b^{2}}-\frac{P^{2}}{b^{4}}}}  \label{FBIcurs}
\end{equation}
$S$\ and $P$\ being the scalar and pseudoscalar field invariants
\begin{equation}
S=\frac{1}{4}\ F_{\mu \nu }F^{\mu \nu
}=\frac{1}{2}(|\mathbf{B}|^{2}-| \mathbf{E}|^{2})
\end{equation}
\begin{equation}
P=\frac{1}{4}\ ^{\ast }F_{\mu \nu }F^{\mu \nu }=\mathbf{E}\cdot \mathbf{B}
\end{equation}
and $^\ast F_{\mu\nu}$ being the dual field tensor (the tensor
resulting from exchanging the roles of ${\bf E}$ and ${-\bf B}$).
When $b\rightarrow \infty $\ then ${\mathcal F}_{\mu \nu
}\rightarrow F_{\mu \nu }$; thus Maxwell equations are recovered.
Equation (\ref{Feq}) means that the field tensor comes from a
four-potential $A_{\mu \;}$: $F_{\mu \nu }=\partial _{\mu }A_{\nu
}-\partial _{\nu }A_{\mu }$. Equation (\ref{Fcurseq}) can be
derived from the Born-Infeld Lagrangian
\begin{equation}
L[A_{\mu }]=-\frac{b^{2}}{4\,\pi }\;\left(
1-\sqrt{1+\frac{2S}{b^{2}}-\frac{ P^{2}}{b^{4}}}\right)
\label{BIL2}
\end{equation}
which goes to Maxwell Lagrangian when $b\rightarrow \infty$. The
unique solutions of Maxwell equations that also solve Born-Infeld
equations (\ref{Feq}, \ref{Fcurseq}) are those having vanishing
$S$\ and $P$, as happens with plane waves.

The simplest solutions for a rectangular waveguide can be obtained
by proposing the field
\begin{equation}
F\ =\ \frac{\partial u(t,x)}{\partial t}\ dt\wedge dy\ +\
\frac{\partial u(t,x)}{\partial x}\ dx\wedge dy  \label{solution1}
\end{equation}
(i.e., $c^{-1}E_{y}=F_{t\,y}=\partial u/\partial
t=-F_{y\,t\;},\;B_{z}=-F_{xy}=-\partial u/\partial x=F_{y\,x},$\
and the rest of the components vanish; the symbol $\wedge$ means
the antisymmetrized tensor product). This field fulfills
Eq.(\ref{Feq}) whatever the function $u(t,x)$\ is (notice that
Eq.(\ref{Feq}) is nothing but $dF=0,$ \ when written in geometric
language). Function $u(t,x)$\ will be determined by
Eq.(\ref{Fcurseq}), together with boundary conditions of the type
$E_{y}(t, x_{boundary})=\partial u/\partial t(t, x_{boundary})=0$.
We will search for an oscillating solution $u(t,x)$. After
obtaining the solution, we will exploit the invariance of
Eqs.(\ref{Feq}, \ref{Fcurseq}) under the Lorentz group, together
with the invariance of the boundary conditions under Lorentz
boosts along the $z$-axis. Thus, by performing a boost
$t\rightarrow \gamma (t-Vc^{-2}z)$, we will obtain a wave
propagating along $z$. The so built solution has the form
\begin{eqnarray}
F\  &=&\ \frac{\partial u(t,x)}{\partial t}|_{t=\gamma
(t-Vc^{-2}z)}\ \gamma \;dt\wedge dy\ \notag \\ &&+\frac{\partial
u(t,x)}{\partial t}|_{t=\gamma (t-Vc^{-2}z)}\ \gamma
Vc^{-2}\;dy\wedge dz\   \notag \\ &&+\ \frac{\partial
u(t,x)}{\partial x}|_{t=\gamma (t-Vc^{-2}z)}\ \;dx\wedge dy
\end{eqnarray}
The $F_{y\,x}$\ component is a magnetic field $B_{z}$\ along the
propagation direction $z$; $F_{t\,y}$\ and $F_{z\,y}$\ are
transversal electric and magnetic fields $c^{-1}E_{y}$ and $B_{x}$
respectively. Therefore, we are building a TE (transverse
electric) mode for a rectangular waveguide. Since the solution
does not depend on the transversal coordinate $y$, then it behaves
like a TE$_{n\,0}$\ mode in Maxwell theory \cite{Jack}. Boundary
conditions for ${\bf E}$ and ${\bf B}$ do not differ from the
usual ones; in fact the continuity of the tangential component of
${\bf E}$ and the normal component of ${\bf B}$ come from
Eq.(\ref{Feq}) that is shared by Maxwell and Born-Infeld theory.
On the other hand, charge and current distributions on the wave
guide surfaces guarantee the boundary conditions required by
Eq.(\ref{Fcurseq}).

When the proposed solution (\ref{solution1}) is substituted in
Eq.(\ref{Fcurseq}) then the following equation for $u(t,x)$\ is
obtained (notice that $P=0$):
\begin{eqnarray}
&&\left[1+\frac{1\,}{b^2}\left( \frac{\partial u}{\partial
x}\right)^2 \right]\ \frac{\partial^2 u}{\partial t^2}\ -\
\frac{2\,}{b^2}\ \frac{
\partial u}{\partial t}\ \frac{\partial u}{\partial x}\ \frac{\partial^2 u
}{\partial t\,\partial x} \notag \\
&&-\left[1-\frac{1\,}{b^2}\left( \frac{\partial u}{\partial
t}\right)^2\right] \ \frac{\partial^2 u}{\partial x^2}\ =\ 0\;,
\label{BIeq}
\end{eqnarray}
which is a wave equation for a scalar field $u(t,x)$ derivable from the
Lagragian $L[u]=(1\,-\,b^{-2}\ \eta ^{\mu \nu }\ \partial _{\mu }\,u\
\partial _{\nu }\,u)^{1/2}$. The Eq.(\ref{BIeq}) is called Born-Infeld
equation; it is integrable \cite{barba,fair} and has a
multi-Hamiltonian structure \cite{arik}. It is hard to find
non-trivial solutions for this equation, especially if one is
looking for solutions accomplishing certain boundary conditions.
Anyway, solutions written in a parametric way have been found
\cite{Pryce1,Pryce2}. In Ref.\cite{Fer} we have developed a
complex method for obtaining solutions of Eq.(\ref{BIeq}) that
uses a Maxwellian solution $u_M$\ as a seed (notice that
Eq.(\ref{BIeq}) becomes $\square\; u_M=0$ when $b\rightarrow
\infty $); the method leads to that solution behaving like the
seed when $b\rightarrow \infty $. Since we have to fit periodic
boundary conditions for the field in the waveguide, we will use
the Maxwellian seed
\begin{equation}
u_{M}(t,x)=\frac{A}{\kappa}\,\cos \kappa ct\;\sin \kappa x
\label{seed}
\end{equation}
(the value of $\kappa$ adjusts the boundary conditions to the
width of the guide along the $x$ direction). Apart from $b$, the
amplitude $A$\ is the only magnitude of the solution having
dimensions of field. Therefore, we expect that the corrections to
the solution coming from the non-linear features of
Eq.(\ref{BIeq}) will depend on the non-dimensional magnitude
$A^{2}b^{-2}$. The Born-Infeld solution associated with the seed
(\ref{seed}) can been worked out by the method explained in
Ref.\cite{Fer}; at the lowest order in $ A^{2}b^{-2}$\ it results
\begin{eqnarray}
u(t,x) &=&\frac{A}{\kappa}\,\Bigg( \cos \left[ \left(
1-\frac{A^{2}}{8\,b^{2}} \right) \;\kappa ct\right]\notag \\ &&
+\frac{A^{2}}{32\,b^{2}}\,\cos \left[ \left( 1-\frac{
A^{2}}{8\,b^{2}}\right) \;3\kappa ct\right] \Bigg)   \notag \\
&&\Bigg( \sin \left[ \left( 1+\frac{A^{2}}{8\,b^{2}}\right)
\;\kappa x\right]
 \label{BIseed}\\ &&+ \frac{A^{2}}{32\,b^{2}}\,\sin \left[ \left(
1+\frac{A^{2}}{8\,b^{2}}\right) \;3\kappa x\right] \Bigg)
+O(A^{4}b^{-4})\notag
\end{eqnarray}
In fact, if the left side of Eq.(\ref{BIeq}) is computed for the
wave (\ref{BIseed}) then an oscillating function of order
$A^{4}b^{-4}$ will be obtained instead of zero. A Lorentz boost
along the $z$-axis turns the solution (\ref{BIseed}) into a
propagating wave. The phase of the temporal factor changes to
$(1-A^{2}/(8b^{2}))\;\kappa\gamma (t-Vc^{-2}z)$ but the phase of
the transversal sector remains invariant. Thus, the frequency and
the wave vector are
\begin{subequations}
\begin{eqnarray}
\omega  &\simeq &\left( 1-\frac{A^{2}}{8\,b^{2}}\right)
\;\kappa\gamma \label{freq} \\ k_{\parallel } &\simeq &\left(
1-\frac{A^{2}}{8\,b^{2}}\right) \;\kappa\gamma Vc^{-2}
\label{long} \\ k_{\perp } &\simeq &\left(
1+\frac{A^{2}}{8\,b^{2}}\right) \;\kappa \label{transv}
\end{eqnarray}
\end{subequations}
Since $\gamma =(1-V^{2}c^{-2})^{-1/2}$\ then we obtain the dispersion
relation
\begin{equation}
\omega ^{2}-c^{2}k_{\parallel }^{2}\simeq \left(
\frac{1-\frac{A^{2}}{ 8\,b^{2}}}{1+\frac{A^{2}}{8\,b^{2}}}\right)
^{2}c^{2}k_{\perp }^{2}\simeq \left(
1-\frac{A^{2}}{2\,b^{2}}\right) c^{2}k_{\perp }^{2}  \label{disp}
\end{equation}
The energy velocity is the velocity $V$ relative to the frame
where the solution is the stationary wave between two boundaries
shown in Eq.(\ref{BIseed}) (in this frame the mean value of the
Poynting vector vanishes). By dividing Eqs. (\ref{long}) and
(\ref{freq}) it is obtained
\begin{equation*}
V=\frac{c^{2}k_{\parallel }}{\omega }
\end{equation*}
then, using the dipersion relation, the energy velocity results
\begin{equation}
\frac{V^2}{c^2}\ =\ 1-\left(1-\frac{A^2}{2\,b^2}\right)\ \frac{c^2
\,k_\perp ^2}{\omega^2}\ +\ O(A^{4}b^{-4}) \label{velocity}
\end{equation}
where $\omega$ is the source frequency and $k_{\perp}$ is
determined by the waveguide width. The Born-Infeld field
configuration in the waveguide is
\begin{eqnarray}
\frac{E_{y}}{\gamma\;A}&=&-\sin (\omega t-k_{\parallel }z)\sin
k_{\perp }x\notag \\ &&+\frac{A^{2} }{32\,b^{2}}\Big[4\sin (\omega
t-k_{\parallel }z)\sin k_{\perp }x\notag \\ &&-\sin (\omega
t-k_{\parallel }z)\sin 3k_{\perp }x \label{field1}\\ &&-3\sin
3(\omega t-k_{\parallel }z)\sin k_{\perp
}x\Big]+O(A^{4}b^{-4})\notag
\end{eqnarray}
\begin{equation}
B_{x}=-V c^{-1}E_{y}  \label{field2}
\end{equation}
\begin{eqnarray}
\frac{B_{z}}{A}&=&-\cos (\omega t-k_{\parallel }z)\cos k_{\perp
}x\notag \\ &&-\frac{A^{2} }{32\,b^{2}}\Big[4\cos (\omega
t-k_{\parallel }z)\cos k_{\perp }x\notag \\ &&+\cos 3(\omega
t-k_{\parallel }z)\cos k_{\perp }x \label{field3}\\ &&+3\cos
(\omega t-k_{\parallel }z)\cos 3k_{\perp
}x\Big]+O(A^{4}b^{-4})\notag
\end{eqnarray}

Eq.(\ref{velocity}) shows that the energy velocity $V$ for
Born-Infeld fields propagating in hollow waveguides increases with
the amplitude $A$, for $A\ll b$. Although the effect is expected
to be very weak, since the Born-Infeld constant $b$ should be a
very large field, the Eq.(\ref{velocity}) shows that the influence
of $A$ on the energy velocity can be magnified by employing a
source wavelength slightly bigger than the transverse dimension of
the waveguide, i.e. when the energy velocity is very small and the
propagating mode is near the cutoff. In particular, the TE$_{10}$
mode will not propagate if the frequency is lower than
$\omega_{cutoff}\simeq(1-b^{-2}A^2/4)\,\pi c/d$, $d$ being the
waveguide width along the $x$ direction; then, the cutoff
frequency decreases with the amplitude. These features are
susceptible of experimental test. Another effect coming from the
non-linearity of the theory is the presence of harmonics into the
field (\ref{field1}-\ref{field3}); so, even in the TE$_{10}$ mode
the waveguide will not be traveled by a unique wavelength, but a
set of harmonics will enter with amplitudes proportional to powers
of $A^{2}b^{-2}$.

\end{document}